\documentclass[12pt,preprint]{aastex}

\slugcomment{Accepted for publication in ApJ}

\shorttitle{Relativistic Jet Interactions with Clouds}
\shortauthors{Choi, Wiita, \& Ryu}

\begin{document}

\title{Hydrodynamic Interactions of Relativistic Extragalactic Jets
       with Dense Clouds}

\author{Eunwoo Choi and Paul J. Wiita}
\affil{Department of Physics and Astronomy, Georgia State University,
       P.O. Box 4106, Atlanta, GA 30302-4106}
\email{echoi@chara.gsu.edu, wiita@chara.gsu.edu}

\and

\author{Dongsu Ryu}
\affil{Department of Astronomy and Space Science,
       Chungnam National University, Daejeon 305-764, Korea}
\email{ryu@canopus.cnu.ac.kr}

\begin{abstract}

We have studied three-dimensional hydrodynamic interactions of
relativistic extragalactic jets with two-phase ambient media.
These jets propagate through a denser homogeneous gas and then impact
clouds with densities $100$ to $1000$ times higher than the initial beam
density.
The deflection angle of the jet is influenced more by the density
contrast of the cloud than by the beam Mach number of the jet.
A relativistic jet with low relativistic beam Mach number can eventually
be slightly bent after it crosses the dense cloud; however, we have not
seen permanently bent structures in the interaction of a high
relativistic beam Mach number jet with a cloud.
The relativistic jet impacts on dense clouds do not necessarily destroy
the clouds completely, and much of the cloud body can survive as a
coherent blob.
This enhancement of cloud durability is partly due to the geometric
influence of the off-axis collisions we consider and also arises from
the lower rate of cloud fragmentation through the Kelvin-Helmholtz
instability for relativistic jets.
To compare our simulations with observed extragalactic radio jets, we
have computed the approximate surface distributions of synchrotron
emission at different viewing angles.
These surface intensity maps show relativistic jets interacting with
clouds can produce synchrotron emission knots similar to structures
observed in many VLBI-scale radio sources.
We find that the synchrotron emission increases steeply at the moment of
impact and the emission peaks right before the jet passes through the
cloud.

\end{abstract}


\section{Introduction}

Relativistic jets emerging from extragalactic sources associated with
active galactic nuclei (AGNs) are the most important means of
transporting energy and mass from AGNs to an external medium over large
distances.
To understand how these relativistic jets interact with an inhomogeneous
external medium containing small, dense gas clouds or clumps has been
recognized as important for a long time.
These interactions may change substantially the direction of
relativistic jet flows, trigger extensive star formation in the shocked
clouds, and possibly explain the basic mechanism behind the morphology
of many extragalactic radio jets \citep{fan74}.

Recent observations have revealed strong evidence of features associated
with changes in jet directions resulting from interactions with
small gas clouds in the narrow-line regions of Seyfert galaxies
\citep[e.g.,][]{mun03}.
Fast outflows of gas observed in the central regions of powerful radio
galaxies can also be caused by such interactions
\citep[e.g.,][]{emo05,mor05}.
The most likely interpretation of fast outflows is that all gas clouds
are not destroyed by the jet; some clouds can severely disrupt the jet
while some clouds are accelerated to the observed high outflow
velocities by the thrust of the jet.
It is argued that despite of the high energies involved in the
interactions, only a few percent of the outflowing gas appears to be
ionized, while the rest of the gas cools and becomes neutral due to
highly efficient cooling near the jet bow shock.

In the context of nonrelativistic hydrodynamic simulations, previous
numerical works were performed to investigate jet interactions with
clouds \citep{deg99,hig99,wan00,sax05} or jets crossing a medium
interface \citep[e.g.,][]{wii90,wii92}, focusing on the effects of the
interactions on the morphology and kinematics of jets.
Others studied shock interactions, focusing on the structure and
evolution of the clouds produced by the interactions in adiabatic cases
\citep{kle94,xu95,pol02} and in radiative cases \citep{mel02,fra04}.
According to \citet{deg99}, simulations with conditions appropriate to
protostellar jets making off-axis collisions with clouds produced a
deflected beam.
The deflection angle tended to decrease with time as the beam slowly
penetrated the cloud and when the jet penetrated most of the cloud, the
deflected beam faded and the jet resumed its original propagation
direction.
\citet{wan00} found the following: powerful extragalactic jets
eventually destroyed the clouds they considered, and these collisions
induced nonaxisymmetric instabilities in the jets; weak jets can be
effectively halted or destroyed by massive clouds; and slow, dense jets
that were bent remained stable for extended times.
Synthetic radio images produced by hydrodynamic simulations for
comparison with observations also supported the hypothesis that these
interactions are responsible for the distorted structures of some radio
jets \citep[e.g.,][]{hig99}.
All those numerical works considered nonrelativistic jet speeds less
than $0.5c$, but the observed apparent superluminal motions of
extragalactic radio sources indicate intrinsic jet speeds up to at least
$0.98c$ \citep{zen97}.
Thus, it is essential to perform relativistic hydrodynamic simulations
of this problem in order to cover the range of true jet speeds.

Since time-dependent numerical simulations of relativistic jets were
first reported \citep{van93,dun94,mar94}, multidimensional relativistic
hydrodynamic simulations have been used as an important method in
understanding of relativistic jets
\citep{mar97,kom98,alo99,ros99,hug02,miz04}.
The morphological and dynamical properties of relativistic jets
propagating through a homogeneous medium were studied by \citet{mar97}
in two dimensions and by \citet{alo99} in three dimensions.
\citet{kom98} investigated the large-scale flows produced by classical
and relativistic jets in a uniform external medium using analytical and
numerical studies.
\citet{hug02} performed in three dimensions a study of the deflection of
relativistic jets by an oblique density gradient and of the precession
of relativistic jets.
They found that fast relativistic jets can be significantly influenced
by an oblique density gradient, showing a rotation of the Mach disk with
the flow bent via a strong oblique internal shock.

In this paper we present results from three-dimensional hydrodynamic
simulations of the interactions of relativistic jets with dense clouds.
We focus on the off-axis collision of the relativistic jet with a steady
spherical cloud.
The main concerns of this study are how the relativistic jets are
influenced by these interactions and how the interaction affects the
evolution of the cloud.
In \S 2 we outline briefly the dynamical problem, while the basic
equations, numerical method and setup we employ are described in \S 3.
In \S 4 we describe the results, and we present a summary and discussion
in \S 5.

\section{Problem Description}

We consider the three-dimensional interactions of relativistic jets with
two-phase ambient media.
These jets propagate through a denser ambient gas and then hit spherical
clouds with densities higher than that of the ambient gas.
The initial ratio of the cloud density, $\rho_c$, to the ambient medium
density, $\rho_a$, and that of the beam density, $\rho_b$, to the
ambient medium density, are respectively defined as
\begin{equation}
\chi\equiv\frac{\rho_c}{\rho_a},~~\eta\equiv\frac{\rho_b}{\rho_a}.
\end{equation}
If we neglect complicating effects including radiative cooling and
gravity and consider only hydrodynamic effects, then this problem can be
relatively simple and depends only on a few hydrodynamic variables, the
Mach number of the jet and the initial density contrasts given in
equation (1).
Any geometric effects, such as different impact zone sizes or cloud
shapes, certainly will make differences in the evolutions of jets and
clouds, but the overall dynamical evolutions should not be very
sensitive to them.
Thus, we will focus on the evolutions of jets and clouds influenced by
the above hydrodynamic effects.

The approximate propagation velocity of the jet through the homogeneous
ambient medium can be obtained by the conservation of the momentum flux
of the beam and ambient medium in the reference frame of the Mach disk
\citep{mar97}.
Assuming pressure equilibrium between the beam and the ambient medium,
the conservation of the momentum flux is
$\rho_bh_b\Gamma_b^{\prime2}v_b^{\prime2} = \rho_ah_a\Gamma_a^{\prime2}v_a^{\prime2}$
with the following relations, $v_b^\prime = (v_b-v_a)/(1-v_bv_a)$,
$\Gamma_b^\prime = \Gamma_b\Gamma_a(1-v_bv_a)$, $v_a^\prime = -v_a$, and
$\Gamma_a^\prime = \Gamma_a$.
Here $h$ is the specific enthalpy and $v^\prime$ and
$\Gamma^\prime$ represent respectively velocity and Lorentz factor
measured in the reference frame of the Mach disk, while $v$ and $\Gamma$
indicate those measured in the rest frame of the ambient medium.
The subscripts $b$ and $a$ stand for the beam and the ambient medium,
respectively.
After substituting for the primed variables in terms of the unprimed
ones, the conservation of the momentum flux is derived to be
\begin{equation}
\rho_bh_b\Gamma_b^2\left(v_b-v_a\right)^2 = \rho_ah_av_a^2.
\end{equation}
Then the one-dimensional jet advance velocity, estimated in the rest
frame of the ambient medium is
\begin{equation}
v_a = \frac{v_b}{\sqrt{1/\eta^*}+1},
\end{equation}
where $\eta^*$ is given by
\begin{equation}
\eta^* = \Gamma_b^2\frac{\rho_bh_b}{\rho_ah_a}.
\end{equation}
In the nonrelativistic limit ($h\rightarrow1$, $\Gamma\rightarrow1$),
$\eta^*$ approaches $\eta$, so that $v_a$ represents the classical jet
advance velocity through the ambient medium, i.e.,
$v_a = v_b/(\sqrt{1/\eta}+1)$.

Based on this jet advance velocity, we define the dynamical timescale
called the {\it beam crossing time} $t_\mathrm{bc}$,
\begin{equation}
t_\mathrm{bc}\equiv\frac{2r_c}{v_a},
\end{equation}
as the time taken for the beam to sweep a distance across the ambient
medium equal to the diameter of a cloud with radius $r_c$.
Since the timescale $t_\mathrm{bc}$ depends only on a single variable
$v_a$ (for fixed cloud radius), it is extremely useful in comparing and
characterizing the dynamical evolutions of both jets and clouds with
different model parameters.

Although we use the beam crossing time as the primary timescale in this
study, it is also interesting to estimate the cloud crushing and cooling
timescales.
The cloud crushing timescale is the time required for the beam to cross
the cloud diameter during the phase of cloud compression, and if $v_a$
is nonrelativistic, this timescale can be approximated as
$t_\mathrm{cc}\sim2\sqrt{\chi}r_c/v_a$ \citep{kle94}.
Clearly, $t_\mathrm{bc}\simeq t_\mathrm{cc}$ in the absence of clouds,
and for dense clouds ($\chi\gg1$), $t_\mathrm{bc}<t_\mathrm{cc}$.
Following \citet{fra04}, the cloud cooling timescale can be roughly
estimated from $t_\mathrm{cool}\sim Cv_a^3/(\chi^{3/2}\rho_c)$, where
the constant
$C = 7.0\times10^{-35}~\mathrm{g}~\mathrm{cm}^{-6}~\mathrm{s}^4$.
With values reasonable for kpc-scale extragalactic situations,
$r_c = 1~\mathrm{kpc}$, $v_a = 0.1c$, $\chi = 100$, and
$\rho_c = 10^2m_\mathrm{H}~\mathrm{cm}^{-3}$, we find that
$t_\mathrm{bc}<t_\mathrm{cc}\sim t_\mathrm{cool}$.
Thus cooling will not be extremely important during the cloud
compression phase for the chosen values.
For fixed density and cloud radius, the cloud cooling timescale becomes
longer compared to the cloud crushing timescale as $v_a$ increases, so
the effect of cooling is somewhat reduced for relativistic jets compared
to nonrelativistic ones.
For parameters more relevant to VLBI-scale jet/cloud collisions,
$r_c = 0.5~\mathrm{pc}$, $v_a = 0.5c$, $\chi = 10^4$, and
$\rho_c = 10^6m_\mathrm{H}~\mathrm{cm}^{-3}$, we have
$t_\mathrm{bc}\sim t_\mathrm{cool}<t_\mathrm{cc}$, so cooling would be
more important in this case.
A more detailed consideration of cooling timescales is beyond the scope
of the current paper.

Three distinct evolutionary stages can be considered in this problem.
There is an initial jet propagation stage where the jet advances through
a homogeneous ambient medium with velocity $v_a$.
Once a jet is launched, a bow shock propagates into the ambient medium;
this is followed by a Mach disk shock in the beam which is quickly
established during this stage.
When the jet strikes the cloud, the jet transmits a shock into the
cloud.
If the cloud/ambient density contrast is sufficiently large and the jet
speed is relatively slow, the speed of the transmitted shock in the
cloud is much slower than that of the bow shock of the jet.
Thus the bow shock entirely encloses the cloud, which leads to the
development of the Kelvin-Helmholtz instability at the cloud surface
\citep[e.g.,][]{kle94}.
The final stage is when the jet passes through the cloud.
In this phase the cloud begins to reexpand just after the jet reaches
the rear edge of the cloud.
At the same time, the jet propagates in the original direction or in a
new direction.

\section{Numerical Simulations}

\subsection{Basic Equations}

The special relativistic hydrodynamic equations are written in a
covariant form \citep[e.g.,][]{lan59,wil03}
\begin{equation}
\partial_\alpha\left(\rho U^\alpha\right) = 0,
\end{equation}
\begin{equation}
\partial_\alpha T^{\alpha\beta} = 0,
\end{equation}
where the energy momentum tensor is given by
\begin{equation}
T^{\alpha\beta} = \left(e+p\right)U^\alpha U^\beta+pg^{\alpha\beta}.
\end{equation}
Here, $\partial_\alpha = \partial/\partial x^\alpha$ is the covariant
derivative with spacetime coordinates $x^\alpha = [t,x_j]$,
$U^\alpha = [\Gamma,\Gamma v_j]$ is the normalized
($U^\alpha U_\alpha = -1$) four-velocity vector, and a metric tensor
$g^{\alpha\beta}$ with a signature $+2$ is used.
The mass density, velocity, internal plus mass energy density, and
pressure in the local rest frame are denoted by $\rho$, $v_j$, $e$, and
$p$, respectively.
Greek indices (e.g., $\alpha$, $\beta$) denote the spacetime components
while Latin indices (e.g., $i$, $j$) indicate the spatial components,
and the speed of light is set to unity ($c\equiv1$) throughout this
paper.

For our numerical purposes, it is convenient to rewrite the covariant
equations (6)--(8) in the index form which gives a hyperbolic system of
conservation equations
\begin{equation}
\frac{\partial D}{\partial t}+\frac{\partial}{\partial x_j}
\left(Dv_j\right) = 0,
\end{equation}
\begin{equation}
\frac{\partial M_i}{\partial t}+\frac{\partial}{\partial x_j}
\left(M_iv_j+p\delta_{ij}\right) = 0,
\end{equation}
\begin{equation}
\frac{\partial E}{\partial t}+\frac{\partial}{\partial x_j}
\left[\left(E+p\right)v_j\right] = 0,
\end{equation}
where the equation of state (EOS) is given by
\begin{equation}
p = \left(\gamma-1\right)\left(e-\rho\right).
\end{equation}
Here, $D$, $M_i$, and $E$ are the mass density, momentum density, and
total energy density in the reference frame, respectively, and $\gamma$
is the adiabatic index.
We note that we restrict ourselves to an ideal gas EOS in this study
\citep[cf.][]{ryu06}.

The quantities in the reference frame are related to those in the local
rest frame via following transformations
\begin{equation}
D = \Gamma\rho,
\end{equation}
\begin{equation}
M_i = \Gamma^2\left(e+p\right)v_i,
\end{equation}
\begin{equation}
E = \Gamma^2\left(e+p\right)-p,
\end{equation}
where the Lorentz factor is given by
\begin{equation}
\Gamma = \frac{1}{\sqrt{1-v^2}}
\end{equation}
with $v^2 = v_x^2+v_y^2+v_z^2$.

If an EOS is assumed, the local sound speed $c_s$ and the specific
enthalpy $h$ are easily derived.
For an ideal gas, they are given by
\begin{equation}
c_s^2 = \frac{1}{h}\frac{\partial p}{\partial\rho}
+\frac{\partial p}{\partial e},~~
h = 1+\frac{\gamma}{\gamma-1}\frac{p}{\rho}.
\end{equation}
A $\gamma$-law gas such as an ideal gas has the local sound speed
limit $c_s\leq\sqrt{\gamma-1}$ and in the ultrarelativistic case
$e\gg\rho$, the local sound speed approaches its limit
(i.e., $c_s\rightarrow\sqrt{\gamma-1}$).

\subsection{Numerical Method and Setup}

The system of equations (9)--(12) can be solved numerically with
explicit finite difference upwind schemes which are based on exact or
approximate Riemann solvers using the characteristic decomposition of
relativistic hydrodynamic conservation equations.
Although the upwind schemes were originally developed for
nonrelativistic hydrodynamics, some schemes have been extended to
special relativistic hydrodynamics retaining the advantages of the
upwind schemes, including high accuracy and robustness.

A multidimensional code for solving the special relativistic
hydrodynamic equations as a hyperbolic system of conservation laws based
on the total variation diminishing (TVD) scheme \citep{har83} was
developed and tested in \citet{cho05}.
The TVD scheme is an explicit Eulerian finite difference upwind scheme
and an extension of the Roe scheme to second-order accuracy in space and
time.
The code uses a new set of conserved quantities, which lead to a new
eigenstructure for special relativistic hydrodynamics and employs an
analytic formula for the calculation of the local rest frame quantities
from the reference frame quantities.
The advantage of our code is that it is simple and fast, and yet it is
accurate and reliable enough.
The performance of the code was demonstrated through several standard
tests, including relativistic shock tubes, a relativistic wall shock,
and a relativistic blast wave, as well as test simulations of the
relativistic version of the Hawley-Zabusky shock and a relativistic
extragalactic jet \citep{cho05}.
For our new simulations, we have parallelized this code using the
message-passing interface (MPI).
The simulations described here typically use $64$ processors on a Linux
cluster.

Table \ref{tab1} lists the initial parameters of the four different
relativistic jet--cloud interaction models we have investigated in this
study.
All models use the adiabatic index $\gamma = 4/3$ and assume
pressure-matched jets and clouds, i.e., $p_b/p_a = p_c/p_a = 1$, where
$p_b$, $p_c$, and $p_a$ are the pressure of the beam, cloud, and ambient
medium, respectively.
We set $c = r_c = \rho_a\equiv1$ in our models, so that all physical
quantities are dimensionless and can be scaled to any specific physical
model (e.g., $t\rightarrow tc/r_c$, $\rho\rightarrow\rho/\rho_a$).
The Newtonian beam Mach number, $\mathcal{M}_b^N\equiv v_b/c_{s,b}$,
where $c_{s,b}$ is the sound speed in the beam, as well as the
relativistic beam Mach number,
$\mathcal{M}_b^R\equiv(\Gamma_b/\Gamma_{s,b})\mathcal{M}_b^N$, where
$\Gamma_{s,b}$ is the Lorentz factor associated with the beam sound
speed are given in Table \ref{tab1}.
As discussed in \citet{kon80}, in the context of relativistic
gasdynamics the relativistic Mach number is the best analog of the
Newtonian one for nonrelativistic flows, so we use this relativistic
beam Mach number to describe physical properties in our models.
The initial density contrast between the beam and the ambient medium is
fixed to $\eta = 0.1$, so that jets strike clouds with densities $100$
to $1000$ times higher than the initial beam density.
In models M1 and M2, the clouds interact with the lower relativistic
beam Mach number jets, where the relativistic effects are less dominant,
with smaller beam velocities and internal energies.
Model M1 is almost identical to model M2 except for the smaller density
ratio of the cloud to the ambient medium.
Models M3 and M4 have been chosen to study the cloud interactions with
the higher relativistic beam Mach number jets, with the more dominant
relativistic effects caused by larger beam velocities and internal
energies.
Again, the initial conditions of model M3 are the same as those of model
M4 except for the smaller density ratio of the cloud to the ambient
medium.

We set up the density gradient of the spherical cloud edge with a
hyperbolic tangent function
\begin{equation}
\rho\left(r\right) = \frac{\rho_c+\rho_a}{2}+\frac{\rho_c-\rho_a}{2}
\tanh\left(\frac{r_c-r}{\Delta r}\right),
\end{equation}
where $r$ is the distance from the center of the cloud and $\Delta r$ is
the scale parameter for the width of density transition
($\Delta r\ll r_c$).
The presence of a true density discontinuity instead of this steep
function would not affect the dynamics of jet--cloud interactions
significantly, but the discontinuous cloud edge is approximated by this
somewhat smoothed density profile to avoid numerical artifacts as the
jet impacts the cloud.
We assume other physical quantities such as pressure and velocity are
constant across the transition width.

The simulations have been performed in the three-dimensional
computational domain with $x = [0,8]$, $y = [0,8]$, and $z = [0,8]$
using a uniform Cartesian grid of $256^3$ cells.
The beam, with a circular cross section of radius $r_b = 1/4$
($8$ cells), is initially located at $(x,y,z) = (0,4,4)$ and propagates
through the ambient medium along the positive $x$-direction.
In order for the relativistic jet to collide off axis with the cloud at
rest, the center of the cloud, with radius $r_c = 1$ ($32$ cells), is
placed at $(x,y,z) = (4,3.5,4)$; hence the relativistic jet hits the
spherical cloud with an impact angle of $30\arcdeg$.
The outflow boundary condition is imposed on all boundaries of the
computational domain except where the inflow boundary condition is used
to maintain the continuous jet.
We were able to assign the relativistic jet only $8$ cells per initial
beam radius and the cloud $32$ cells per initial cloud radius due to the
limitation of computational resources.
This resolution is less than that of previously reported two-dimensional
works which are related to this problem.
Thus our three-dimensional simulations may not be fully converged and some
quantities to be described may change if three-dimensional simulations
with much higher resolutions are performed in future studies; however,
our tests of the code do indicate that simulations with this level of
resolution should be reasonably accurate \citep{cho05}.

\section{Results}

\subsection{Morphology and Dynamics}

\begin{figure}
\begin{center}
\epsscale{0.95}
\plotone{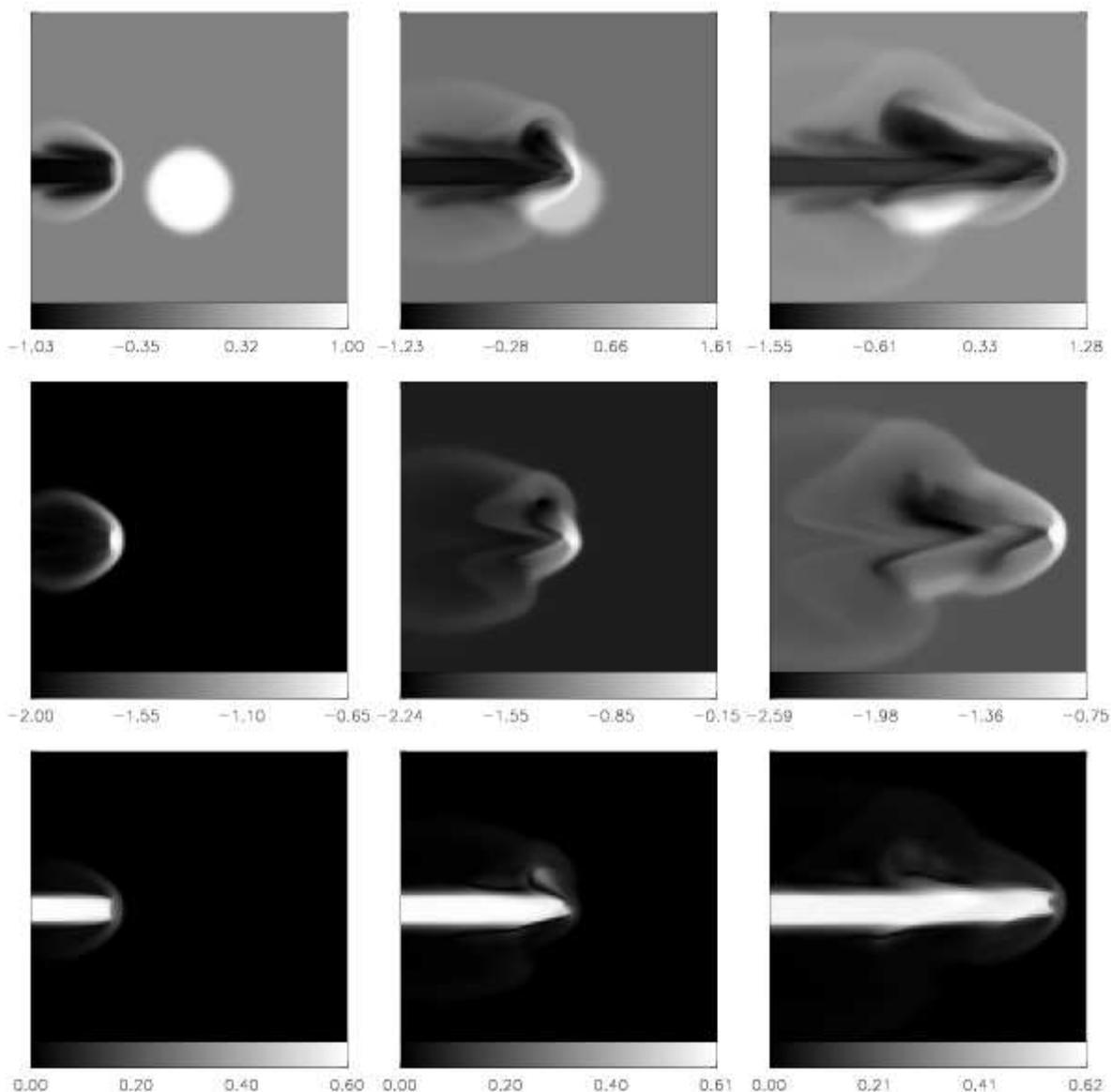}
\end{center}
\caption{(a) Gray-scale images of density, pressure, and Lorentz factor
(top to bottom) for model M1 at three different times,
$t/t_\mathrm{bc} = 1$, $2.5$, and $4$ (left to right).
The image scales are logarithmic for $\rho$ and $p$ but the square-root
of the logarithm for $\Gamma$ so as to enhance visibility of
intermediate values; the images show the $x-y$ plane with $z = 4$ in the
three-dimensional computational domain.
(b) Same as Fig. \ref{fig1}(a) except for model M2 at
$t/t_\mathrm{bc} = 1$, $3.5$, and $6$.
(c) Same as Fig. \ref{fig1}(a) except for model M3 at
$t/t_\mathrm{bc} = 1$, $2.5$, and $4$.
(d) Same as Fig. \ref{fig1}(a) except for model M4 at
$t/t_\mathrm{bc} = 1$, $3$, and $5$.}
\label{fig1}
\end{figure}

The gray-scale images in Figures \ref{fig1}(a)--(d) show the distinct
evolutionary phases of models M1--M4, respectively.
These images show the $x-y$ plane with $z = 4$ in the three-dimensional
computational domain.
In each of these figures the top to bottom panels represent density,
pressure, and Lorentz factor, respectively (in logarithmic scales) while
the left to right panels represent evolutionary stages shown at three
different times, $t = t_\mathrm{bc}$,
$(t_\mathrm{bc}+t_\mathrm{end})/2$, and $t_\mathrm{end}$.

The early stages of the relativistic jet propagation through the uniform
ambient medium until the jet is about to collide the cloud
($t/t_\mathrm{bc}\sim1$) are basically similar to those found in earlier
simulations \citep[e.g.,][]{mar97,alo99}.
Several key features are clear from the left panels of Figures
\ref{fig1}(a)--(d).
In all the models a bow shock that separates the jet from the external
medium is driven, the beam itself is terminated by a Mach disk (terminal
shock) where the beam kinetic energy is converted into its internal
energy, and shocked jet material flows backward along the contact
discontinuity (working surface) into a cocoon.
There is no difference between models M1 and M2 and between models M3
and M4 at this stage because of the same initial conditions of the jets
and the same ambient media properties for these two pairs of
simulations.

The relativistic beam Mach number of the jet is associated with the
shape of the bow shock.
In models M1 and M2, the lower relativistic beam Mach number jets, with
a lower propagation velocity ($v_a\sim0.42$) and internal energy, have
bow shocks with narrower conical shapes, and the Mach disk is quite
close to the bow shock.
This conical shape of the bow shock tends to be broader as the
relativistic beam Mach number of the jet increases, as seen for models
M3 and M4; these higher relativistic beam Mach number jets, with a
higher propagation velocity ($v_a\sim0.78$) and internal energy, also
have the Mach disk standing off farther from the bow shock.
The shapes of the bow shocks are also connected with the sizes of the
impact cross section when the jets begin to interact with the cloud.
The low relativistic beam Mach number jets in models M1 and M2 feature
relatively thick cocoons while the high relativistic beam Mach number
jets in models M3 and M4 have thin cocoons.
This dependence of the cocoon morphology on the relativistic beam Mach
number is consistent with previous results \citep[see e.g.,][]{mar97}.
Although the structural differences in the jet head and the cocoon are
evident by this early stage of the evolution, the internal structures
within the beam and backflows are not dominant and are barely
distinguishable at this stage.

In every model, the relativistic jet begins to partially deflect as a
direct response to its interactions with the clouds.
This is seen in the middle column of panels of Figures \ref{fig1}(a)--(d);
seen most clearly in the middle bottom panels are fast streams coming
from the Mach disk at significant angles with respect to the jet axis.
These deflection features are stronger in models M2 and M4, which have
higher ratio of the density of the cloud to that of the ambient medium
($\chi = 100$).
The deflection angles of the portion of the post-Mach shock flows with
respect to the beam propagation axis are very time-dependent.
In our models these angles peak when the jets cross over approximately
half the clouds (at $t/t_\mathrm{bc}\sim2.5$, $3.5$, $2.5$, and $3$ for
models M1--M4, respectively).
For the comparable dynamical times, models M2 and M4, both with
$\chi = 100$ but having different beam Mach numbers, show
$80\arcdeg-90\arcdeg$ deflection angles, while models M1 and M3 with the
same beam Mach numbers as the models M2 and M4, respectively, but with
$\chi = 10$, show smaller deflection angles of about $45\arcdeg$.
This indicates that the deflection angle is more strongly influenced by
the density contrast, $\chi$, than by the beam Mach number of the jet.
For an off-axis collision there are weak deflection features on the
other side of the jet axis, where the deflection of the outflow from the
beam is significantly suppressed by the dense cloud.
That suppression leads to the production of a strong oblique shock
within the beam.
As seen in the figures, the oblique shocks are quite strong in models M2
and M4, but in models M1 and M3 there are only relatively weak oblique
shocks in the beam.
Comparing at this stage models M1 and M3 with models M2 and M4, we note
that the bow shocks enclose less of the cloud in models M1 and M3
because of their lower density contrast, $\chi$.
That implies quicker penetration of the clouds by these jets, so the
strengths of the oblique shocks in these beams are reduced.

Some additional properties of the simulations at this stage are shown in
Figures \ref{fig2} and \ref{fig3}.
Figure \ref{fig2} illustrates one-dimensional flow structures of
density, pressure, and Lorentz factor along the beam propagation axis
for models M1 and M3 at the same epoch as in Figures \ref{fig1}(a) and (c).
In both models there are spikes in the density and pressure associated
with the impact by the incident jets, while there is little change in
beam Lorentz factor.
Figure \ref{fig3} shows the images of the logarithm of the Lorentz
factor projected at the viewing angle of $0\arcdeg$ for models M2 and
M4, at $t/t_\mathrm{bc} = 3.5$ and $3$, respectively.
These projection images clearly show the anisotropic distribution and
directions of the deflected gas induced by the jet.
This gas is an admixture of jet and cloud material, but only a small
fraction of the cloud gas is shown in these projection images since the
mean cloud velocity computed in each component (refer to \S 4.2) is
$\langle v_i\rangle\lesssim0.01$ and $0.06$ for models M2 and M4,
respectively, at the same epoch as in Figure \ref{fig3}.
This implies that this deflected gas consists predominantly of jet
material although small amount of cloud material is entrained in these
deflected structures.
This presence of deflected gas accelerating toward a terminal velocity
strongly suggests that such deflected and accelerated gas is responsible
for at least some of the outflowing gas observed in the vicinity of AGNs.

Once the jet passes through the cloud, it begins to accelerate, causing
a change in the shape of the bow shock.
As visible in the right panels of Figures \ref{fig1}(a)--(d) shown when
the jet head nearly reaches the boundary of the computational cube
(at $t/t_\mathrm{bc} = 4$, $6$, $4$, and $5$ for models M1--M4,
respectively), the shape of the bow shocks changes more clearly in the
low relativistic beam Mach number jets than in the high relativistic
beam Mach number jets.
That reflects the fact that the acceleration of the jets is somewhat
faster in low relativistic beam Mach number jets.
That reacceleration occurs in essentially the original propagation
direction or in a somewhat new direction.
In our simulations there is a trend for the flow of the jet to be bent
more when a lower relativistic beam Mach number jet interacts with a
denser cloud, with the least bending seen for model M3 and the most for
model M2.
We see in the right panels of Figure \ref{fig1}(b) that the beam is bent
by about $10\arcdeg$ with respect to the original jet axis.
The bent jet still remains stable and collimated over the several
dynamical times we could follow its development.

After the jet head passes the cloud, the amount of strongly deflected
gas gradually reduces and the oblique shocks continue to develop in the
beam.
These oblique shocks are unlikely to play a major role in slowing the
jets because we do not find any significant deceleration features during
this stage.
Although a significant portion of the momentum flux of the jets is
transfered to the deflected gas and the cloud through the collision
events, the jets in our simulations are still stable and well collimated
over several dynamical times after collisions even if the jet is bent.
This stable, collimated condition is quantitatively apparent in the flow
structures of density, pressure, and Lorentz factor shown in Figure
\ref{fig2} at $t/t_\mathrm{bc} = 4$ (corresponding to the dashed lines)
for models M1 and M3, respectively.
There are only slight fluctuations in the flow structures at this late
stage.

In comparing our simulations with hydrodynamic simulations of
nonrelativistic jet--cloud interactions \citep{deg99,hig99,wan00} we can
only note some fairly basic similarities and differences between our
relativistic models and the roughly corresponding nonrelativistic models.
The lack of good overlap between the $\mathcal{M}_b^R$ values for the
relativistic jets and the standard Mach number for the nonrelativistic
jets as well as differences between cloud size to jet-width ratios
considered here and in that earlier nonrelativistic work prevents us
from making quantitative comparisons.
Relativistic jets interacting with dense clouds certainly do show general
morphological features such as deflections of some gas and bent
structures of jets similar to those found in some of the nonrelativistic
jet--cloud interactions.
The slower relativistic jet shows a bent structure after interaction,
which is similar to that found in nonrelativistic simulations involving
``weak'' jets while the faster relativistic jet effectively plows through
the clouds.
Higher power nonrelativistic jets also can plow through, and apparently
completely destroy, clouds.
However, some major differences arise because of the larger propagation
velocity of the relativistic jets.
Because of this, moderately light ($\eta = 0.1$) relativistic jets are
not effectively decelerated and disrupted by the dense ($\chi = 100$)
clouds, whereas nonrelativistic jets assaulting clouds of similar density
ratios typically are disrupted.
Our relativistic jets are rather reaccelerated in either a slightly new or
essentially the original direction after their interactions with clouds.
The large propagation velocity also suppresses the development of
hydrodynamic instabilities in the jets, so that the jets still remain
stable and collimated even after the jets smash into much denser clouds.
In addition, as discussed in \S 4.2, clouds impaled by relativistic jets
also appear to survive somewhat better than do those hit by strong
nonrelativistic jets.

\subsection{Cloud Evolutions}

Although previous studies of jet interactions with clouds mainly
emphasized the dynamical and morphological features of the jet itself,
it is also important to follow the evolution of the clouds during and
after the off-axis collisions with relativistic jets.
One key reason for investigating the fate of the clouds is that the
leftover cloud material is a strong candidate for star formation regions
in the vicinity of AGNs \citep[e.g.,][]{ree89,gop01}.
The cloud is expected to undergo a somewhat different evolution in our
case compared with the evolution of the clouds struck by the
nonrelativistic planar shocks considered in earlier work
\citep[e.g.,][]{kle94,xu95,mel02,fra04}.

In order to describe the evolution of a cloud quantitatively, we
introduce a conserved variable $f$ called a Lagrangian tracer
\citep[e.g.,][]{jon96} which is updated along with the primitive
hydrodynamic variables in our simulations.
The evolution of the Lagrangian tracer is followed by the conservation
equation
\begin{equation}
\frac{\partial Df}{\partial t}+\frac{\partial}{\partial x_j}
\left(Dfv_j\right) = 0.
\end{equation}
Since the above conservation equation is almost identical to the mass
conservation equation (9), it is separately solved using the same TVD
routine adopted for solving the mass conservation equation.
Initially the tracer variable is set to unity ($f_c = 1$) inside the
cloud while the variable is set to zero ($f_c = 0$) everywhere outside
the cloud, so that the density of cloud material is given as
$D_c = Df_c$ (i.e., $\rho_c = \rho f_c$) for a given tracer $f_c$ in any
zone.
Then the total mass of the cloud is computed by the integration over the
entire volume $V$,
\begin{equation}
m_c = \int_V\rho_cdV,
\end{equation}
where $dV = dxdydz$.
This enables us to compute the several useful mass-weighted quantities
such as the mean square radius of the cloud and the mean velocity of the
cloud,
\begin{equation}
\langle r_i^2\rangle = \frac{1}{m_c}\int_V\rho_cr_i^2dV,
\end{equation}
\begin{equation}
\langle v_i\rangle = \frac{1}{m_c}\int_V\rho_cv_idV.
\end{equation}
The index $i$ given above represents each spatial component.
Another useful mass-weighted quantity is the mean thermal energy inside
the cloud $\langle e_\mathrm{th}\rangle$.
This is also computed using the same volume integration given above.

We show in Figure \ref{fig4} the volume-rendering images of cloud
density for model M4 at three different times, $t/t_\mathrm{bc} = 1$,
$3$, and $5$.
As a direct consequence of the impact on the cloud by the jet the cloud
develops a cavity in the cloud body as shown in the figure.
The cloud cavity continues to grow until the jet completely penetrates
the cloud, elongating the cloud material outside the cavity along the
bow shock of the jet.
Unlike the cases studied earlier where a cloud interacts with a
plane-parallel shock \citep{kle94,xu95}, the cloud material is not
completely destroyed by the impact of the jet.
Some cloud mass is carried into the deflected material of the jet,
eroding the cloud body, but much of the cloud mass remains in a large,
coherent blob for at least a few beam crossing times.
This enhancement of the cloud durability is apparently primarily due to
the geometric influence of an off-axis collision.
Computational resource limits prescribe that we can accurately
investigate the clouds for only a few beam crossing times, which is less
than the many dynamical times for which it would be desirable to follow
their evolutions.

Figure \ref{fig5} shows for every model the time evolutions of the root
mean square radius of cloud, the mean cloud velocity, and the mean
thermal energy of the cloud.
In every model the clouds remain in the initial root mean square radius
$\langle r_i^2\rangle^{1/2} = 0.44$ until $t/t_\mathrm{bc}\sim1.5$.
When the jet hits the cloud, the cloud is first crushed in the
$x$-direction, along which the jet propagates, and then it begins to
expand beyond its initial size.
The initial compressions in the $y$- and $z$-directions are very small
and the cloud soon gradually expands in both these transverse
directions.
By the end of these simulations the root mean square radii of the clouds
have expanded to about $1.5-2$ times their initial values.

After $t/t_\mathrm{bc}\sim1.5$ the high pressure inside the cloud
generated by the incident jet causes the entire cloud to accelerate.
Unsurprisingly, the acceleration is faster in the $x$-direction for the
faster jets in models M3 and M4 and for the lighter clouds in models
M1 and M3.
The mean velocity of the clouds peaks at values between $0.01$ and
$0.15$ in the $x$-direction and $0.005$ and $0.05$ in the $y$-direction.
Note that the mean velocity of the clouds in the $z$-direction is zero
because of symmetry in this direction.
The maximum velocity of the cloud is always rather modest even if the
incident jet has a relativistic speed, though if we had considered less
massive clouds they obviously could have been accelerated to higher
speeds.

As we expect, the mean thermal energy of the cloud increases while the
jet strikes the cloud.
The maximum mean thermal energy of the cloud reaches about $5-15$ times
its initial value, depending upon the model.
In each model the peaks of the mean thermal energy inside the cloud and
the mean velocity of the cloud take place nearly at the same time.
Note that at this point the cloud reexpands after the cloud reaches the
maximum compression in the $x$-direction.

As mentioned earlier, the jet interaction with the cloud shows that the
beam penetrates through the cloud body, which may begin a fragmentation
process.
A strong shear layer developing at the cloud boundary as a result of the
interaction with the jet may lead to Kelvin-Helmholtz instabilities
which enable the disrupted cloud body to fragment.
So eventually the gas cloud might be broken into small pieces.
However, the Kelvin-Helmholtz instability becomes inefficient if the
density contrast of two slipping fluids is large or if the flow is
supersonic \citep{cha61}, so we may not see rapid fragmentation in the
clouds.
Although the fragmentation timescale is difficult to estimate, our
simulations show no significant cloud fragmentation by
$t/t_\mathrm{bc}\sim4-6$.
This indicates that the high density contrast between clouds and beams
and supersonic velocity of the clouds induced by the relativistic jets
do indeed lower the growth rate of the Kelvin-Helmholtz instabilities.

\subsection{Synchrotron Emission}

Propagating relativistic jets produce nonthermal radio (synchrotron)
emission which originates from relativistic high-energy particles
accelerated at the shock front.
\citet{jon99} and \citet{tre01} calculated the synchrotron emission in
extragalactic jets by explicitly calculating the acceleration of
electrons at shocks and following the evolution of magnetic field.
However, they assumed nonrelativistic jets, and hence the emissivity
needs to be further examined using relativistic jets.
To compute the synchrotron emission from relativistic jets, other
relativistic hydrodynamic simulations have worked with a simpler
approximation \citep{gom97,kom97,mio97,alo00}.
Using this same simple model we now calculate the synchrotron emission
in our simulations in order to estimate how the relativistic jet
interaction with a cloud would appear in emission as a extragalactic
radio source.
We make the usual assumptions that the jet is optically thin and only
the jet material radiates.
Thus, in order to separate the jet material from the ambient medium and
the cloud, we include an additional tracer variable $f_b$ (see \S 4.2)
which is initially set to unity inside the jet ($f_b = 1$) and zero
everywhere outside the jet ($f_b = 0$).

The relativistic high-energy electrons responsible for the synchrotron
emission are assumed to have a power-law energy distribution.
Given the spectral index $\alpha$, the high-energy particle number
density $N_0$, and the magnetic field intensity $B$, the synchrotron
emissivity at frequency $\nu$ is then approximated by the power-law
distribution \citep[see e.g.,][]{mio97}
\begin{equation}
j_\nu\propto N_0B^{\alpha+1}\nu^{-\alpha}.
\end{equation}
The high-energy particle number density, $N_0$, is assumed to be
proportional to the relativistic electron energy density, $u_e$, from
the integration of the power-law energy distribution over some energy
range, and $u_e$ is also taken to be proportional to the hydrodynamic
pressure.
Then we have $N_0\propto u_e\propto p$.
Assuming that there is an equipartition of the magnetic field energy
density $u_B$ and the relativistic electron energy density
($u_B = u_e$), then $u_B\propto p$.
This leads to $B\propto u_B^{1/2}\propto p^{1/2}$.
Therefore, equation (23) becomes
\begin{equation}
j_\nu\propto p^{\left(\alpha+3\right)/2}\nu^{-\alpha}.
\end{equation}
This equation shows that the local thermal pressure approximately
reflects the local synchrotron emissivity.
We have used $\alpha = 0.75$ in our calculation.
By integrating the synchrotron emissivity along the line of sight $L$ at
a viewing angle $\theta$, we can compute the synchrotron intensity on
the surface projected onto the line of sight at the viewing angle
\begin{equation}
I_\nu = \int_L\mathcal{D}^2j_\nu dL,
\end{equation}
where the Doppler boosting factor is given by
\begin{equation}
\mathcal{D} = \frac{1}{\Gamma\left(1-v\cos\theta\right)}.
\end{equation}
Other relativistic effects, including light aberration and time
dilation, have not been included in this calculation, as we assume that
these effects are negligible.

Figure \ref{fig6} shows the synchrotron intensity maps of models M1 and
M2 at the viewing angles $90\arcdeg$, $45\arcdeg$, and $0\arcdeg$.
These maps are shown at $t/t_\mathrm{bc} = 2.5$ for M1 and $3.5$ for M2
when the jet is colliding with the cloud.
The peak intensity in this figure varies with the models and the angles
of view.
Doppler boosting has a little effect on the emission of the jet at the
viewing angle $90\arcdeg$, so that the observed emission is very closely
related to the intrinsic emissivity in this case.
At smaller viewing angles (e.g., $45\arcdeg$ and $0\arcdeg$), however,
the emission morphology is determined to a large degree by Doppler
boosting.
The synchrotron emission is dominated by the bright hotspot, which takes
the form of the compact emission knot in VLBI radio maps.
Although the beam and the deflected material show only weak emission
features, there is a faint secondary spot seen from deflected material
in model M2.

The time evolution of the total synchrotron intensity for models M1--M4
at the viewing angles of $90\arcdeg$, $45\arcdeg$, and $0\arcdeg$ are
shown in Figure \ref{fig7}.
The total synchrotron intensity computed here is in arbitrary units.
There are significant quantitative differences among the models, but the
intensity curves show qualitatively the same trends.
The total synchrotron intensity is much amplified at smaller viewing
angles of $45\arcdeg$ and $0\arcdeg$ because Doppler boosting plays a
role in the amplification of the intensity in these cases.
As expected, the passage of the jet over a cloud enhances the
synchrotron intensity; there are high amplitude bumps in the intensity
curves during the interactions.
The total intensity steeply increases at the moment of the impact by the
jet, and then gradually increases until the jet crosses over the cloud.
This tells us that the compression of the plasma in this region produces
higher synchrotron emission in this approximation where it is tied to
the pressure.
The peak synchrotron intensity occurs shortly after the jet passes
through the entire cloud, and after that the intensity falls off slowly
because the compression is weaker.

Although we have not computed the thermal X-ray emission in detail, we
can briefly discuss it.
Since the free-free emission (bremsstrahlung) is proportional to
$\rho^2$, the total X-ray luminosity due to thermal bremsstrahlung
is most sensitive to the density of gas, provided that the gas is, or
becomes, hot enough to emit X-rays.
The relativistic jet itself is not expected to emit thermal X-rays
because of its low density, and only in some cases does the synchrotron
spectrum extend far enough to produce nonthermal X-ray emission
\citep{har06}.
The dense cloud is very unlikely to start out hot enough to emit X-rays.
However, during the jet--cloud interaction the density and pressure of
the cloud become so high that the total X-ray emission may be larger
inside the cloud than elsewhere, for example, in the bow shock of the jet.
An estimate of the increase in the total X-ray luminosity of the cloud
is given by
$L_x/L_{x,0}\simeq(\rho_c/\rho_{c,0})^2(T_c/T_{c,0})^{1/2}(V_c/V_{c,0})$,
where $L_x$ is the total X-ray luminosity of the cloud, $T_c$ is the
mean cloud temperature, $V_c$ is the mean cloud volume, and the
subscript $0$ represents the initial (preshocked) value.
If we simply assume an ideal gas so $T_c\propto p_c/\rho_c$, we have
$L_x/L_{x,0}\sim(\rho_c/\rho_{c,0})^{3/2}(p_c/p_{c,0})^{1/2}(V_c/V_{c,0})$,
allowing us to estimate the total X-ray luminosity of the cloud with
respect to its preshocked X-ray luminosity.
In model M4, for example, $\rho_c/\rho_{c,0}\approx3$,
$p_c/p_{c,0}\approx12$, and $V_c/V_{c,0}\approx1$ at $t/t_\mathrm{bc} = 3$
(as can be roughly estimated from Figs. \ref{fig4} and \ref{fig5}),
so that $L_x/L_{x,0}\sim18$.
Thus, the shocked cloud could possibly be a important source of thermal
X-rays depending upon the various physical parameters such as the
incident jet velocity, the cloud density, and, most importantly, the
initial cloud temperature, which is not explicitly specified in our
scaled models.
However any thermal significant X-ray luminosity should subside rapidly
after the interaction as the cloud is diffusive and quickly attains
equilibrium with the postshocked ambient pressure.

\section{Summary and Discussion}

We have performed three-dimensional relativistic hydrodynamic
simulations to study relativistic jet interactions with dense clouds,
focusing on the influence of special relativistic effects.
We have investigated clouds struck by both low and high relativistic
beam Mach number jets which have less and more dominant relativistic
effects, respectively, and have compared our results to the extent
possible with nonrelativistic simulations which have been published
previously.
We also have studied the evolution of the assaulted clouds and have
estimated the synchrotron emission from the relativistic jets
interacting with the clouds.

In our models, the partial deflections of the jets due to the
interactions with clouds are seen more clearly when denser clouds are
involved, and the deflection angle is more strongly influenced by the
density contrast of the cloud to the ambient medium than by the beam
Mach number of the jet.
The streams of deflected gas from the jet induced by the interactions
move outward much faster compared to nonrelativistic models.
If our models can be generalized, this suggests that the relativistic
jet--cloud interactions are an effective mechanism of producing at least
some of the outflows observed in the vicinity of AGNs
\citep[e.g.,][]{emo05,mor05}.
After the relativistic jets interact with the dense clouds, we find that
the slower relativistic jets can be bent by modest angles and that these
bent jets still remain stable and collimated over fairly extended
timescales.
This trend is similar to the results from nonrelativistic simulations.

The impact of the jet erodes the cloud, but much of the cloud mass
survives as a large coherent body rather than being completely
destroyed.
This enhancement of the cloud durability compared to interactions with
planar shocks appears to be primarily due to the geometric influence of
an off-axis collision.
Compared to head-on collisions, off-axis collisions damage the cloud
less, increasing the chance of survival of a large portion of the cloud.
Another likely reason for the enhancement of the cloud durability is
that the rate of cloud fragmentation through Kelvin-Helmholtz
instabilities is lowered since the relativistic flows reduce the growth
rate of the instabilities compared to similar off-axis blows by
nonrelativistic jets.
This leftover tenacious cloud material could be a candidate for strong
star formation region in the vicinity of AGNs, particularly when the
cooling timescales are sufficiently short.

The synchrotron intensity ``maps'' show that at the jet impact on a
cloud, the synchrotron emission comes dominantly from a bright hotspot
which could correspond to the form of the compact emission knots seen in
many VLBI radio maps.
Although the emission from the deflected jet material is relatively
weak, there is a secondary synchrotron spot visible from this deflected
material.
This emission feature may represent some of the distorted emission seen
in many VLBI radio maps.
The passage of the jet over a cloud significantly enhances the total
synchrotron intensity of the jet.
We find that the synchrotron emission is steeply enhanced shortly after
the jet hits the cloud, but the emission peaks right before the jet
passes through the cloud.
The next big step in performing these calculations would be to include
magnetic fields.
Such relativistic magnetohydrodynamical simulations would allow for
better estimates of synchrotron emission and would allow the examination
of the polarization of emission arising from more complicated shock
structures.
Such polarization structures could be a useful diagnostic of the
dynamics.

Although most astrophysical simulations based on relativistic
hydrodynamics, including this study, have assumed the ideal EOS, it is
well known that the ideal EOS is correct only if the gas is assumed to
be entirely nonrelativistic ($\gamma = 5/3$) or ultrarelativistic
($\gamma = 4/3$).
If a local transition between nonrelativistic gas and relativistic gas
is involved, the ideal EOS will produce incorrect results in that
regime.
Recently, \citet{ryu06} have studied this issue of the EOS in numerical
relativistic hydrodynamics and propose a new EOS which is simple and yet
approximates closely the EOS of a perfect gas in the relativistic
regime, having an accuracy in enthalpy better than $0.8\%$.
Future numerical simulations using this new EOS should produce even
better results concerning the problem of relativistic jet interactions
with clouds.

\acknowledgments

We thank the anonymous referee for several suggestions which improved
the presentation in this paper.
The simulations were performed on the Linux Biocluster at Georgia State
University, and we are grateful for the allocation of substantial time
on this cluster.
EC and PJW have been supported in part by a subcontract to GSU from NSF
grant AST 0507529 to the University of Washington.
DR has been supported in part by the KOSEF grant R01-2004-000-10005-0.

\clearpage

\begin{deluxetable}{ccccccccc}
\tablewidth{0pt}
\tablecaption{Simulation parameters
\label{tab1}}
\tablehead{
\colhead{Model} &
\colhead{$\chi$} &
\colhead{$\eta$} &
\colhead{$v_b$} &
\colhead{$\Gamma_b$} &
\colhead{$\mathcal{M}_b^N$} &
\colhead{$\mathcal{M}_b^R$} &
\colhead{$t_\mathrm{bc}$} &
\colhead{$t_\mathrm{end}$}}
\startdata
 M1 & 10  & 0.1 & 0.9  & 2.29 & 2.92 & 6.36 & 4.86 & 4$t_\mathrm{bc}$ \\
 M2 & 100 & 0.1 & 0.9  & 2.29 & 2.92 & 6.36 & 4.86 & 6$t_\mathrm{bc}$ \\
 M3 & 10  & 0.1 & 0.99 & 7.09 & 1.92 & 11.6 & 2.50 & 4$t_\mathrm{bc}$ \\
 M4 & 100 & 0.1 & 0.99 & 7.09 & 1.92 & 11.6 & 2.50 & 5$t_\mathrm{bc}$ \\
\enddata
\tablecomments{Here $\chi$ is the ratio of the cloud density to the
ambient medium density, $\eta$ is the ratio of the beam density to the
ambient medium density, $v_b$ is the initial beam velocity, $\Gamma_b$
is the beam Lorentz factor, $\mathcal{M}_b^N$ is the Newtonian beam Mach
number, $\mathcal{M}_b^R$ is the relativistic beam Mach number,
$t_\mathrm{bc}$ is the beam crossing time, and $t_\mathrm{end}$ is the
time at which the simulation is ended.}
\end{deluxetable}

\clearpage

\clearpage

\begin{figure}
\begin{center}
\epsscale{0.95}
\plotone{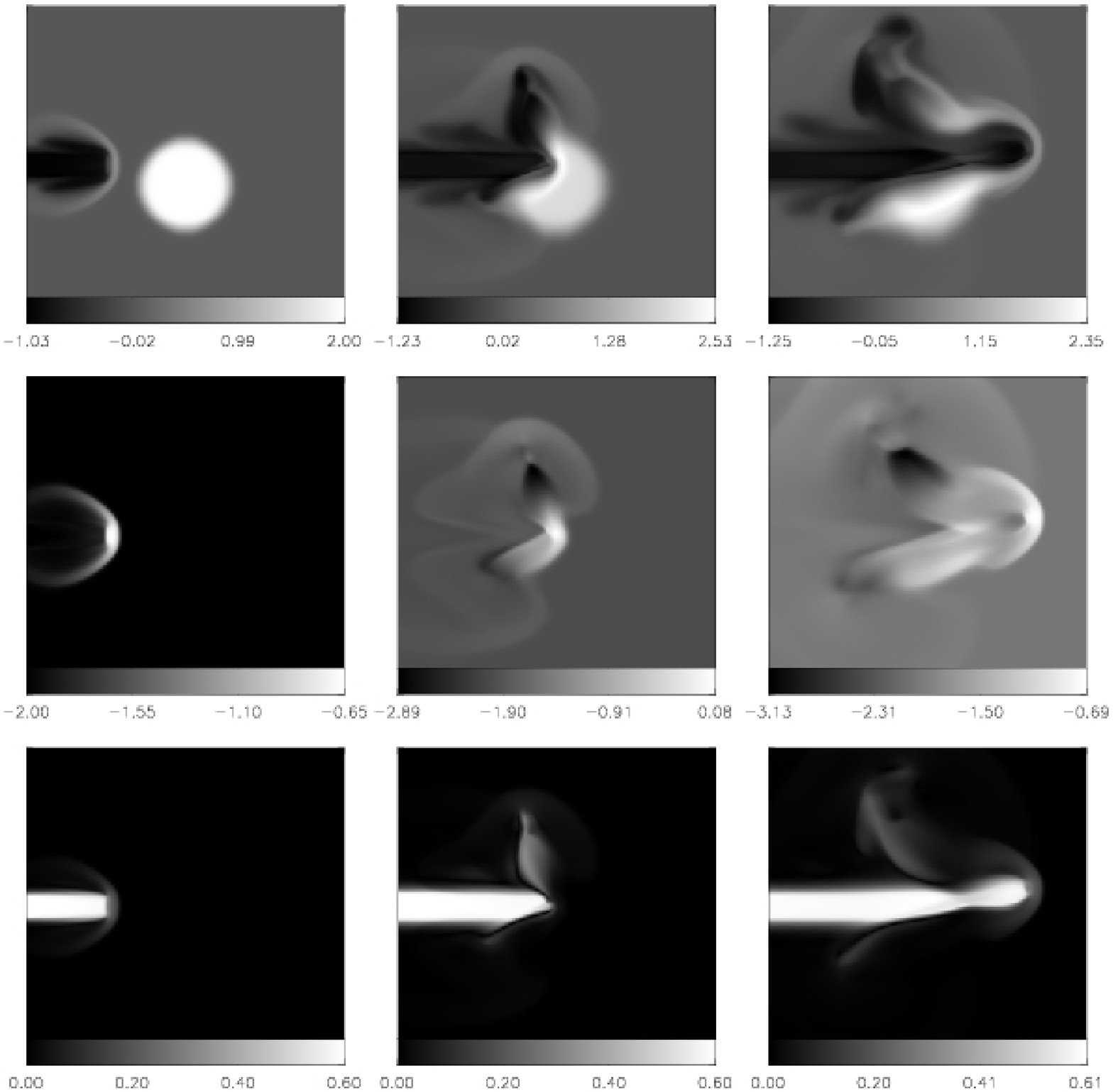}
\end{center}
\end{figure}

\clearpage

\begin{figure}
\begin{center}
\epsscale{0.95}
\plotone{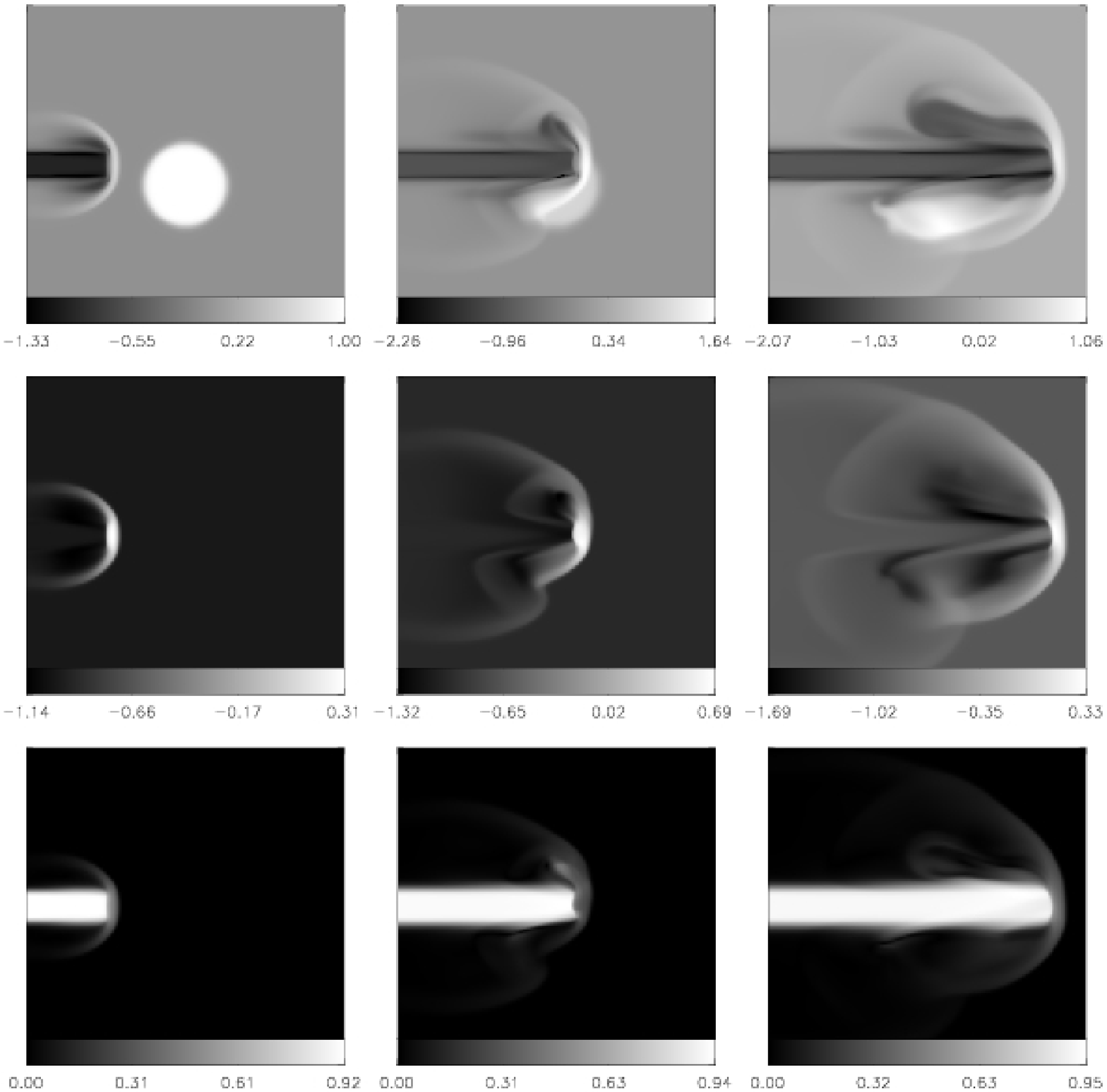}
\end{center}
\end{figure}

\clearpage

\begin{figure}
\begin{center}
\epsscale{0.95}
\plotone{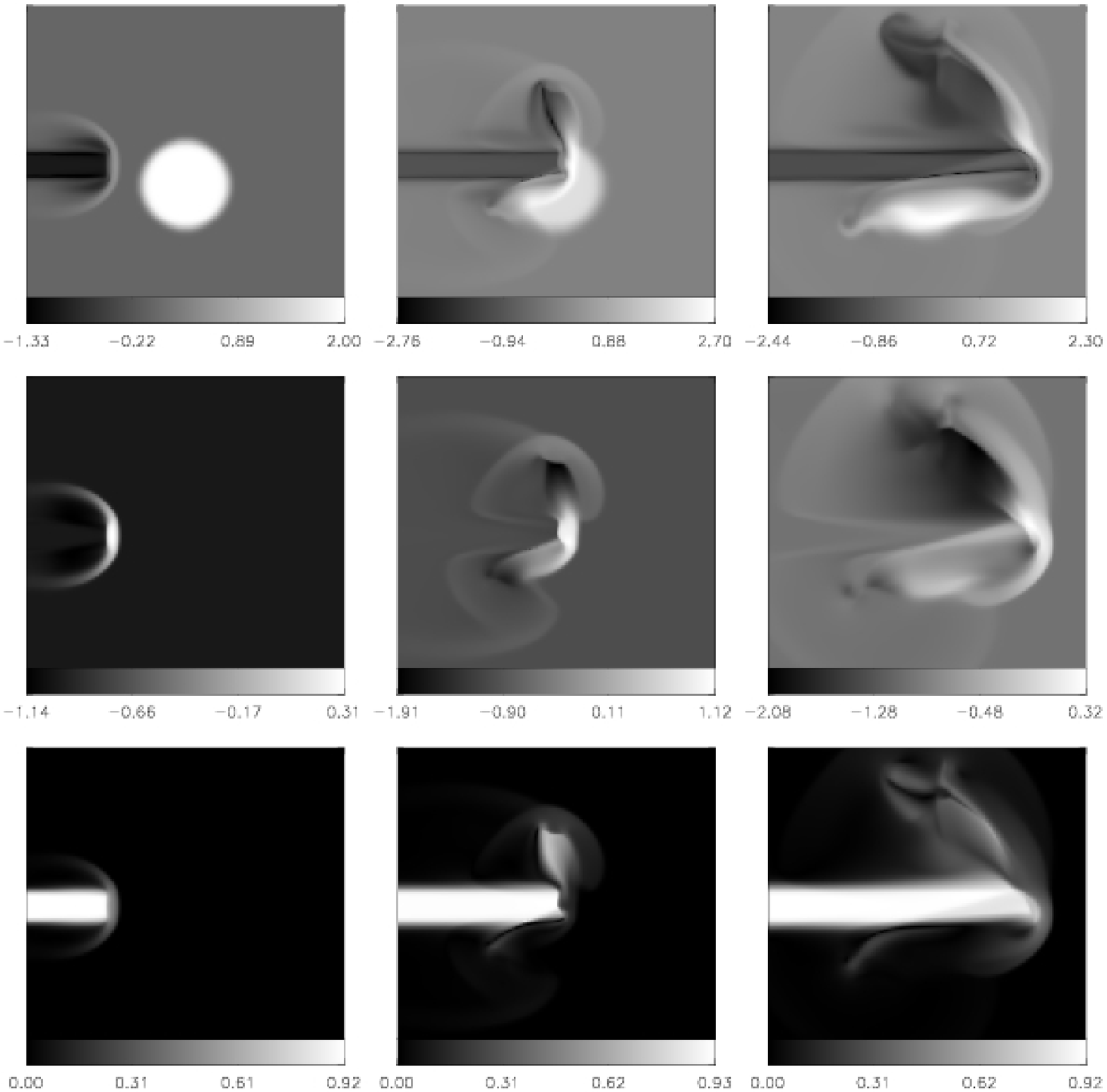}
\end{center}
\end{figure}

\clearpage

\begin{figure}
\begin{center}
\epsscale{1.0}
\plotone{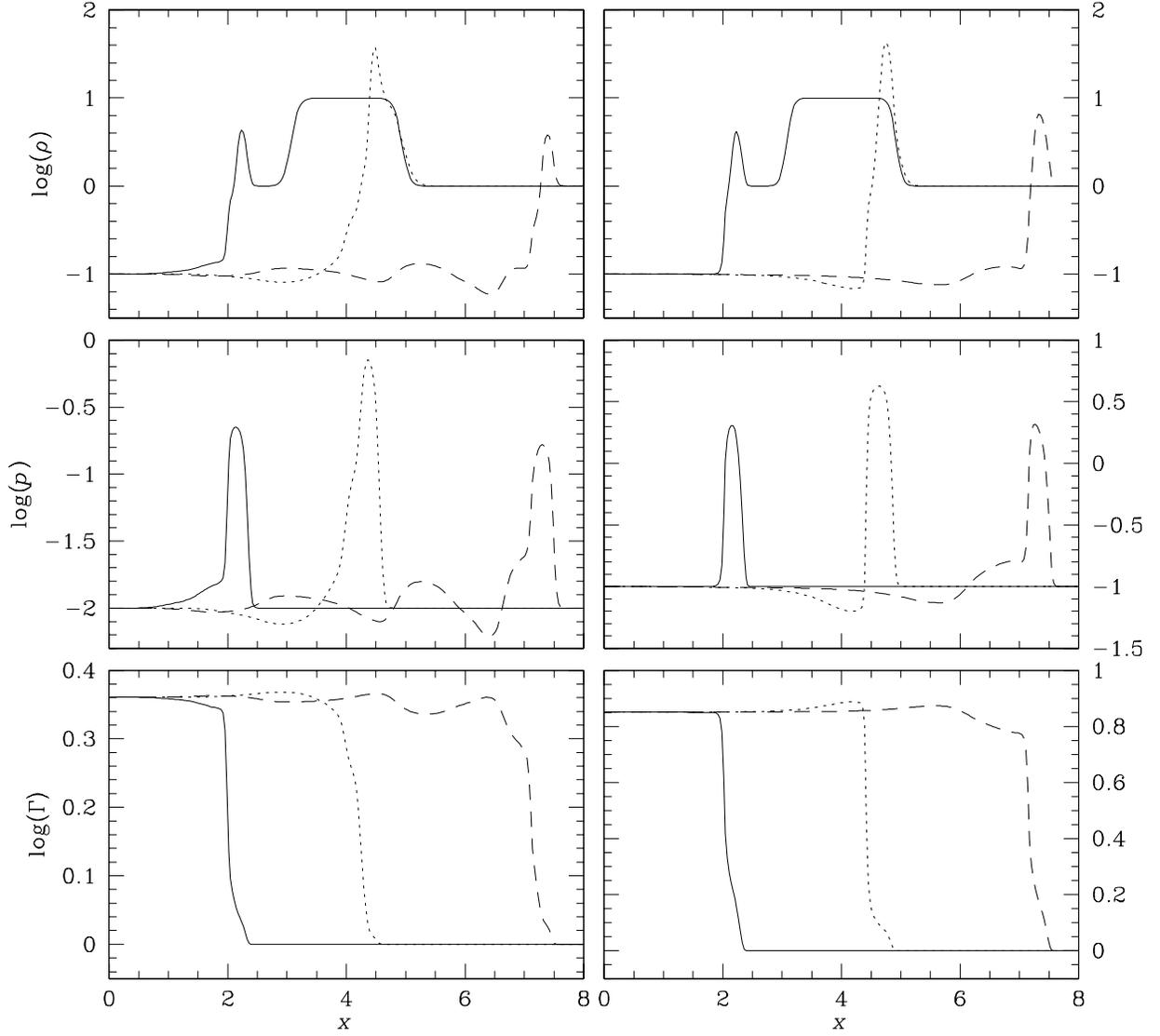}
\end{center}
\caption{Distributions of density, pressure, and Lorentz factor along
the beam propagation axis with $y = z = 4$ for models M1 (left) and M3
(right) at the same epoch as in Figs. \ref{fig1}(a) and (c).
The solid lines correspond to $t/t_\mathrm{bc} = 1$, dotted lines
represent profiles at $t/t_\mathrm{bc} = 2.5$, and dashed lines
illustrate quantities at $t/t_\mathrm{bc} = 4$.}
\label{fig2}
\end{figure}

\clearpage

\begin{figure}
\begin{center}
\epsscale{0.5}
\plotone{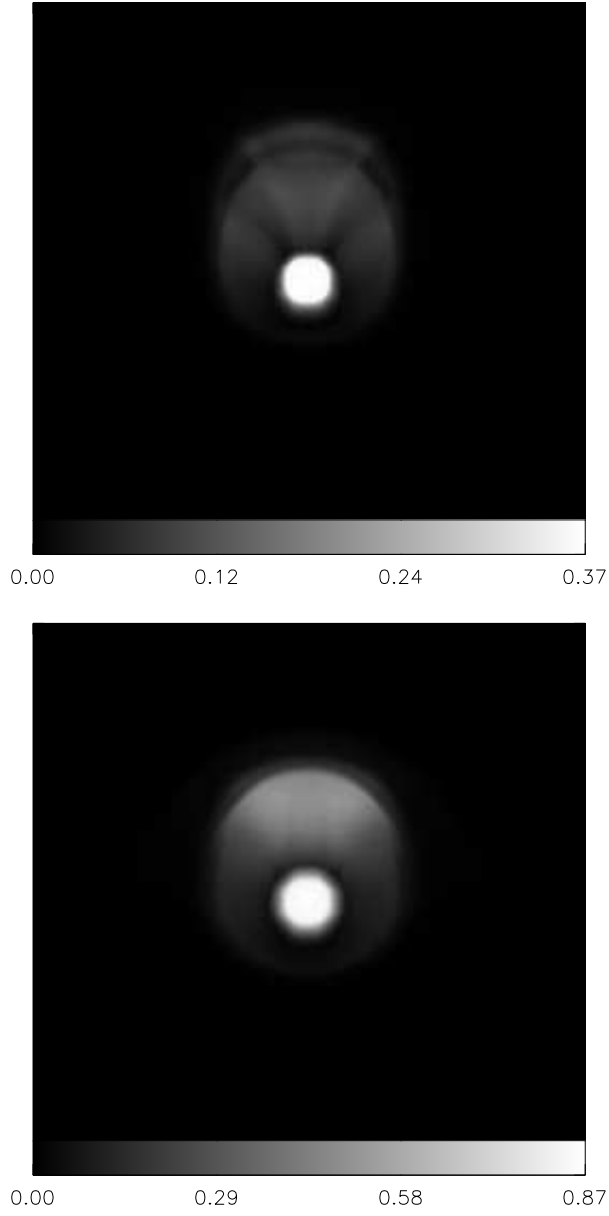}
\end{center}
\caption{Projection images of Lorentz factors at viewing angles of
$0\arcdeg$ for models M2 (top) and M4 (bottom) at
$t/t_\mathrm{bc} = 3.5$ (M2) and $3$ (M4).
The image scales are logarithmic and the images are projected on the
$y-z$ plane in the three-dimensional computational domain.}
\label{fig3}
\end{figure}

\clearpage

\begin{figure}
\begin{center}
\epsscale{0.35}
\plotone{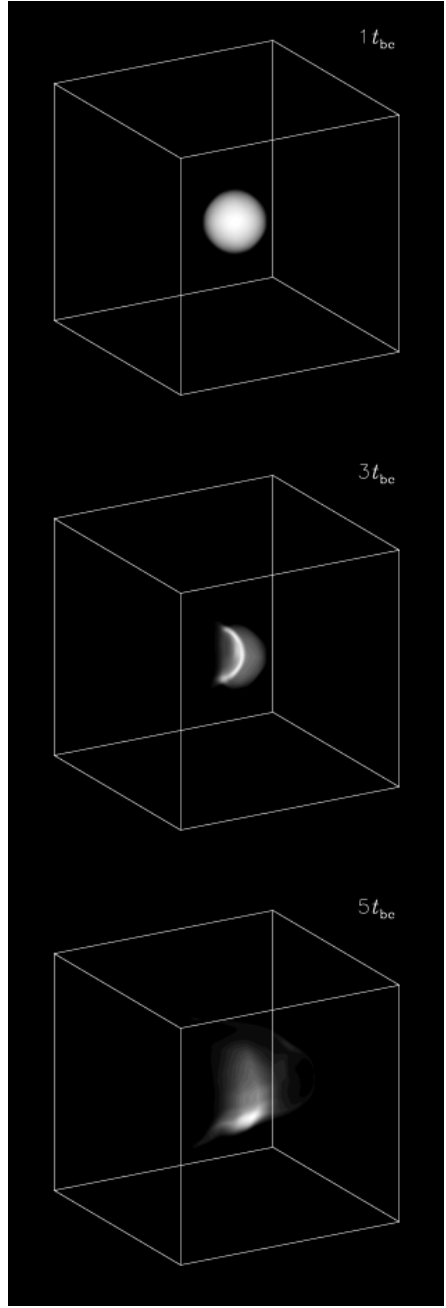}
\end{center}
\caption{Volume-rendering images of cloud density for model M4 at three
different times, $t/t_\mathrm{bc} = 1$, $3$, and $5$.
The image scales are linear and the viewing area is rotated $20\arcdeg$
about the $x$-axis and $30\arcdeg$ about the $z$-axis.
Black represents the lowest values which are $\sim0$ at each epoch and
white the highest values which are $\sim100$ at $t/t_\mathrm{bc} = 1$,
$\sim464$ at $t/t_\mathrm{bc} = 3$, and $\sim229$ at
$t/t_\mathrm{bc} = 5$.}
\label{fig4}
\end{figure}

\clearpage

\begin{figure}
\begin{center}
\epsscale{1.0}
\plotone{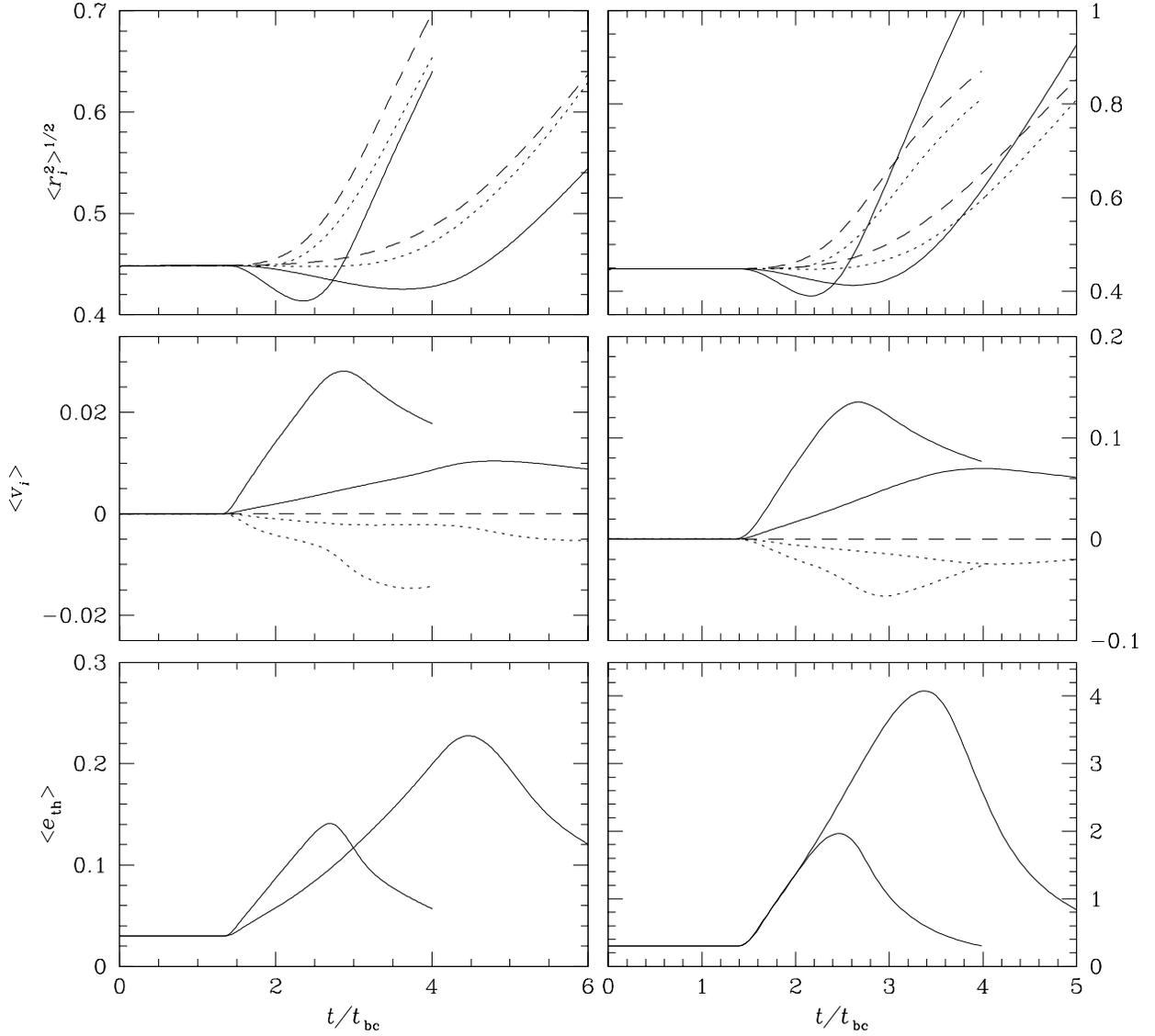}
\end{center}
\caption{Time evolutions of the root mean square radius of the cloud,
the mean cloud velocity, and the mean thermal energy of the cloud for
models M1 (curves ending at $t/t_\mathrm{bc} = 4$) and M2 (curves ending
at $t/t_\mathrm{bc} = 6$) in the left panels and for models M3 (curves
ending at $t/t_\mathrm{bc} = 4$) and M4 (curves ending at
$t/t_\mathrm{bc} = 5$) in the right panels.
The solid, dotted, and dashed lines in the root mean square radius and
the mean velocity panels represent the $x$-, $y$-, and
$z$-components, respectively.}
\label{fig5}
\end{figure}

\clearpage

\begin{figure}
\begin{center}
\epsscale{0.7}
\plotone{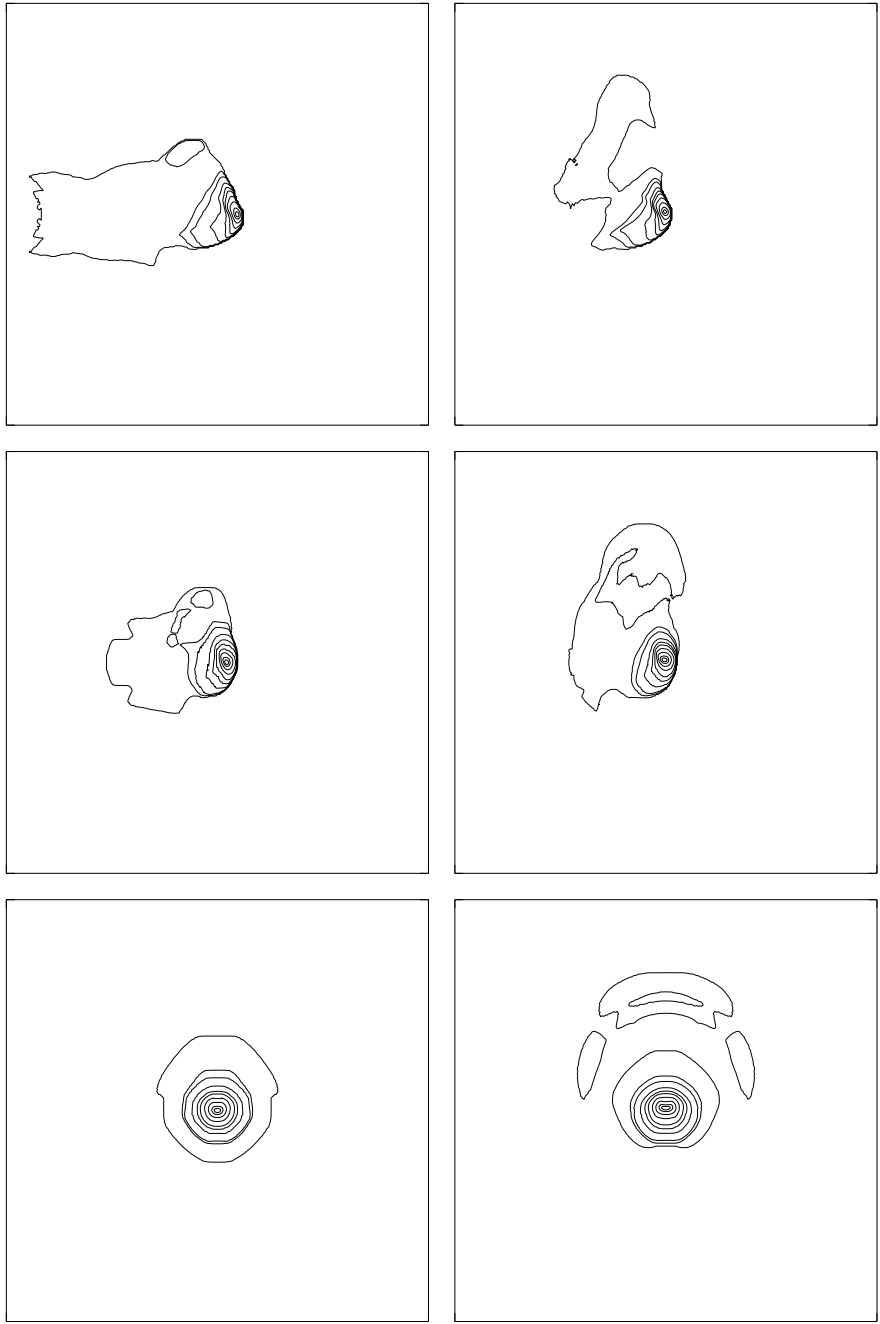}
\end{center}
\caption{Contour maps of synchrotron intensity at the viewing angles of
$90\arcdeg$, $45\arcdeg$, and $0\arcdeg$ (top to bottom) for models M1
(left) and M2 (right) at $t/t_\mathrm{bc} = 2.5$ (M1) and $3.5$ (M2).
Maximum synchrotron intensities are $0.17$ ($90\arcdeg$), $0.13$
($45\arcdeg$), and $0.27$ ($0\arcdeg$) for model M1 and $0.57$
($90\arcdeg$), $0.35$ ($45\arcdeg$), and $0.55$ ($0\arcdeg$) for model
M2, and the contour levels are $0.1\%$, $0.5\%$, $1\%$, $3\%$, $6\%$,
$10\%$, $20\%$, $40\%$, $70\%$, and $90\%$ of the maximum synchrotron
intensity.}
\label{fig6}
\end{figure}

\clearpage

\begin{figure}
\begin{center}
\epsscale{1.0}
\plotone{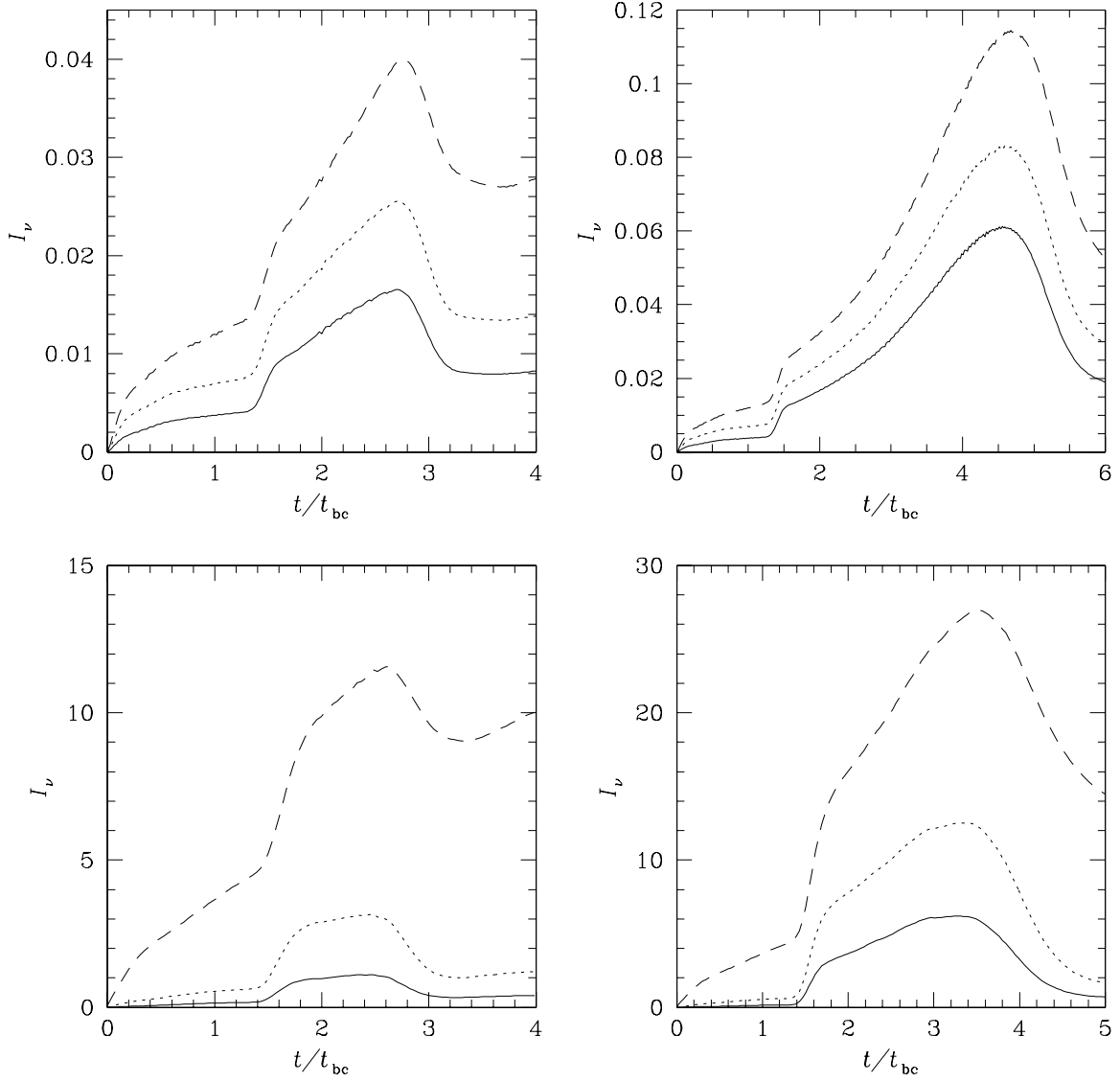}
\end{center}
\caption{Total synchrotron intensity curves for models M1 (top left), M2
(top right), M3 (bottom left), and M4 (bottom right).
The solid, dotted, and dashed lines correspond to the viewing angles of
$90\arcdeg$, $45\arcdeg$, and $0\arcdeg$, respectively.}
\label{fig7}
\end{figure}


\begin{thebibliography}{}

\bibitem[Aloy et al.(2000)]{alo00}
 Aloy, M. A., G\'omez, J. L., Ib\'a\~nez, J. M., Mart\'\i, J. M., \&
 M\"uller, E. 2000, \apj, 528, L85
\bibitem[Aloy et al.(1999)]{alo99}
 Aloy, M. A., Ib\'a\~nez, J. M., Mart\'\i, J. M., G\'omez, J. L., \&
 M\"uller, E. 1999, \apj, 523, L125
\bibitem[Chandrasekhar(1961)]{cha61}
 Chandrasekhar, S. 1961, Hydrodynamic and Hydromagnetic Stability
 (New York: Oxford Univ. Press)
\bibitem[Choi \& Ryu(2005)]{cho05}
 Choi, E., \& Ryu, D. 2005, \na, 11, 116
\bibitem[de Gouveia Dal Pino(1999)]{deg99}
 de Gouveia Dal Pino, E. M. 1999, \apj, 526, 862
\bibitem[Duncan \& Hughes(1994)]{dun94}
 Duncan, G. C., \& Hughes, P. A. 1994, \apj, 436, L119
\bibitem[Emonts et al.(2005)]{emo05}
 Emonts, B. H. C., Morganti, R., Tadhunter, C. N., Oosterloo, T. A.,
 Holt, J., \& van der Hulst, J. M. 2005, \mnras, 362, 931
\bibitem[Fanaroff \& Riley(1974)]{fan74}
 Fanaroff, B. L., \& Riley, J. M. 1974, \mnras, 167, 31P
\bibitem[Fragile et al.(2004)]{fra04}
 Fragile, P. C., Murray, S. D., Anninos, P., \& van Breugel, W.
 2004, \apj, 604, 74
\bibitem[G\'omez et al.(1997)]{gom97}
 G\'omez, J. L., Mart\'\i, J. M., Marscher, A. P., Ib\'a\~nez, J. M., \&
 Alberdi, A. 1997, \apj, 482, L33
\bibitem[Gopal-Krishna \& Wiita(2001)]{gop01}
 Gopal-Krishna, \& Wiita, P. J. 2001, \apj, 560, L115
\bibitem[Harris \& Krawczynski(2006)]{har06}
 Harris, D. E., \& Krawczynski, H. 2006, \araa, 44, 463
\bibitem[Harten(1983)]{har83}
 Harten, A. 1983, J. Comput. Phys., 49, 357
\bibitem[Higgins et al.(1999)]{hig99}
 Higgins, S. W., O'Brien, T. J., \& Dunlop, J. S. 1999, \mnras, 309, 273
\bibitem[Hughes et al.(2002)]{hug02}
 Hughes, P. A., Miller, M. A., \& Duncan, G. C. 2002, \apj, 572, 713
\bibitem[Jones et al.(1999)]{jon99}
 Jones, T. W., Ryu, D., \& Engel, A. 1999, \apj, 512, 105
\bibitem[Jones et al.(1996)]{jon96}
 Jones, T. W., Ryu, D., \& Tregillis, I. L. 1996, \apj, 473, 365
\bibitem[Klein et al.(1994)]{kle94}
 Klein, R. I., McKee, C. F., \& Colella, P. 1994, \apj, 420, 213
\bibitem[Komissarov \& Falle(1997)]{kom97}
 Komissarov, S. S., \& Falle, S. A. E. G. 1997, \mnras, 288, 833
\bibitem[Komissarov \& Falle(1998)]{kom98}
 Komissarov, S. S., \& Falle, S. A. E. G. 1998, \mnras, 297, 1087
\bibitem[K\"onigl(1980)]{kon80}
 K\"onigl, A. 1980, Phys. Fluids, 23, 1083
\bibitem[Landau \& Lifshitz(1959)]{lan59}
 Landau, L. D., \& Lifshitz, E. M. 1959, Fluid Mechanics
 (London: Pergamon Press)
\bibitem[Mart\'\i~et al.(1997)]{mar97}
 Mart\'\i, J. M., M\"uller, E., Font, J. A., Ib\'a\~nez, J. M., \&
 Marquina, A. 1997, \apj, 479, 151
\bibitem[Mart\'\i~et al.(1994)]{mar94}
 Mart\'\i, J. M., M\"uller, E., \& Ib\'a\~nez, J. M. 1994, \aap, 281, L9
\bibitem[Mellema et al.(2002)]{mel02}
 Mellema, G., Kurk, J. D., \& R\"ottgering, H. J. A.
 2002, \aap, 395, L13
\bibitem[Mioduszewski et al.(1997)]{mio97}
 Mioduszewski, A. J., Hughes, P. A., \& Duncan, G. C.
 1997, \apj, 476, 649
\bibitem[Mizuta et al.(2004)]{miz04}
 Mizuta, A., Yamada, S., \& Takabe, H. 2004, \apj, 606, 804
\bibitem[Morganti et al.(2005)]{mor05}
 Morganti, R., Oosterloo, T. A., Tadhunter, C. N., van Moorsel, G., \&
 Emonts, B. 2005, \aap, 439, 521
\bibitem[Mundell et al.(2003)]{mun03}
 Mundell, C. G., Wrobel, J. M., Pedlar, A., \& Gallimore, J. F.
 2003, \apj, 583, 192
\bibitem[Poludnenko et al.(2002)]{pol02}
 Poludnenko, A. Y., Frank, A., \& Blackman, E. G. 2002, \apj, 576, 832
\bibitem[Rees(1989)]{ree89}
 Rees, M. J. 1989, \mnras, 239, 1P
\bibitem[Rosen et al.(1999)]{ros99}
 Rosen, A., Hughes, P. A., Duncan, G. C., \& Hardee, P. E.
 1999, \apj, 516, 729
\bibitem[Ryu et al.(2006)]{ryu06}
 Ryu, D., Chattopadhyay, I., \& Choi, E. 2006, \apjs, 166, 410
\bibitem[Saxton et al.(2005)]{sax05}
 Saxton, C. J., Bicknell, G. V., Sutherland, R. S., \& Midgley, S.
 2005, \mnras, 359, 781
\bibitem[Tregillis et al.(2001)]{tre01}
 Tregillis, I. L., Jones, T. W., \& Ryu, D. 2001, \apj, 557, 475
\bibitem[van Putten(1993)]{van93}
 van Putten, M. H. P. M. 1993, \apj, 408, L21
\bibitem[Wang et al.(2000)]{wan00}
 Wang, Z., Wiita, P. J., \& Hooda, J. S. 2000, \apj, 534, 201
\bibitem[Wiita \& Norman(1992)]{wii92}
 Wiita, P. J., \& Norman, M. L. 1992, \apj, 385, 478
\bibitem[Wiita et al.(1990)]{wii90}
 Wiita, P. J., Rosen, A., \& Norman, M. L. 1990, \apj, 350, 545
\bibitem[Wilson \& Mathews(2003)]{wil03}
 Wilson, J. R., \& Mathews, G. J. 2003, Relativistic Numerical
 Hydrodynamics (Cambridge: Cambridge Univ. Press)
\bibitem[Xu \& Stone(1995)]{xu95}
 Xu, J., \& Stone, J. M. 1995, \apj, 454, 172
\bibitem[Zensus(1997)]{zen97}
 Zensus, J. A. 1997, \araa, 35, 607

\end{thebibliography}
\end{document}